\documentclass[aps,prl,twocolumn,superscriptaddress]{revtex4-1}
\usepackage{graphicx}
\usepackage{amsmath,amssymb}
\usepackage{multirow}
\usepackage{dsfont}

\usepackage{color}

\begin{document}

\title{
Quantum criticality of quasi one--dimensional topological Anderson insulators
}

\author{Alexander Altland}

\affiliation{Institut f\"ur Theoretische Physik, Universit\"at zu K\"oln,
Z\"ulpicher Stra\ss e 77, 50937 K\"oln, Germany}

\author{Dmitry Bagrets}

\affiliation{Institut f\"ur Theoretische Physik, Universit\"at zu K\"oln,
Z\"ulpicher Stra\ss e 77, 50937 K\"oln, Germany}

\author{Lars Fritz}

\affiliation{Institut f\"ur Theoretische Physik, Universit\"at zu K\"oln,
Z\"ulpicher Stra\ss e 77, 50937 K\"oln, Germany}

\affiliation{Institute for Theoretical Physics, Utrecht University, Leuvenlaan 4, 3584 CE Utrecht, Netherlands}

\author{Alex Kamenev}

\affiliation{W. I. Fine Theoretical Physics Institute and School of Physics and Astronomy, University
of Minnesota, Minneapolis, MN 55455, USA}

\author{Hanno Schmiedt}

\affiliation{Institut f\"ur Theoretische Physik, Universit\"at zu K\"oln,
Z\"ulpicher Stra\ss e 77, 50937 K\"oln, Germany}

\date{\today}

\begin{abstract}
We present an analytic theory of quantum criticality in the quasi one-dimensional topological Anderson insulators of class $\mathrm{AIII}$ and
$\mathrm{BDI}$. We describe the systems  in terms of two parameters $(g,\chi)$
representing localization and topological properties, respectively. Surfaces
of half-integer valued $\chi$ define phase boundaries between distinct
topological sectors. Upon increasing system size, the two parameters exhibit
flow similar to the celebrated two parameter flow describing the class
$\mathrm{A}$ quantum Hall insulator. However, unlike the quantum Hall
system, an exact analytical description of the entire phase diagram can be given. 
We check the quantitative validity of our theory by comparison to
numerical transfer matrix computations.
\end{abstract}

\maketitle

\begin{figure}[b]
\includegraphics[width=8.6cm]{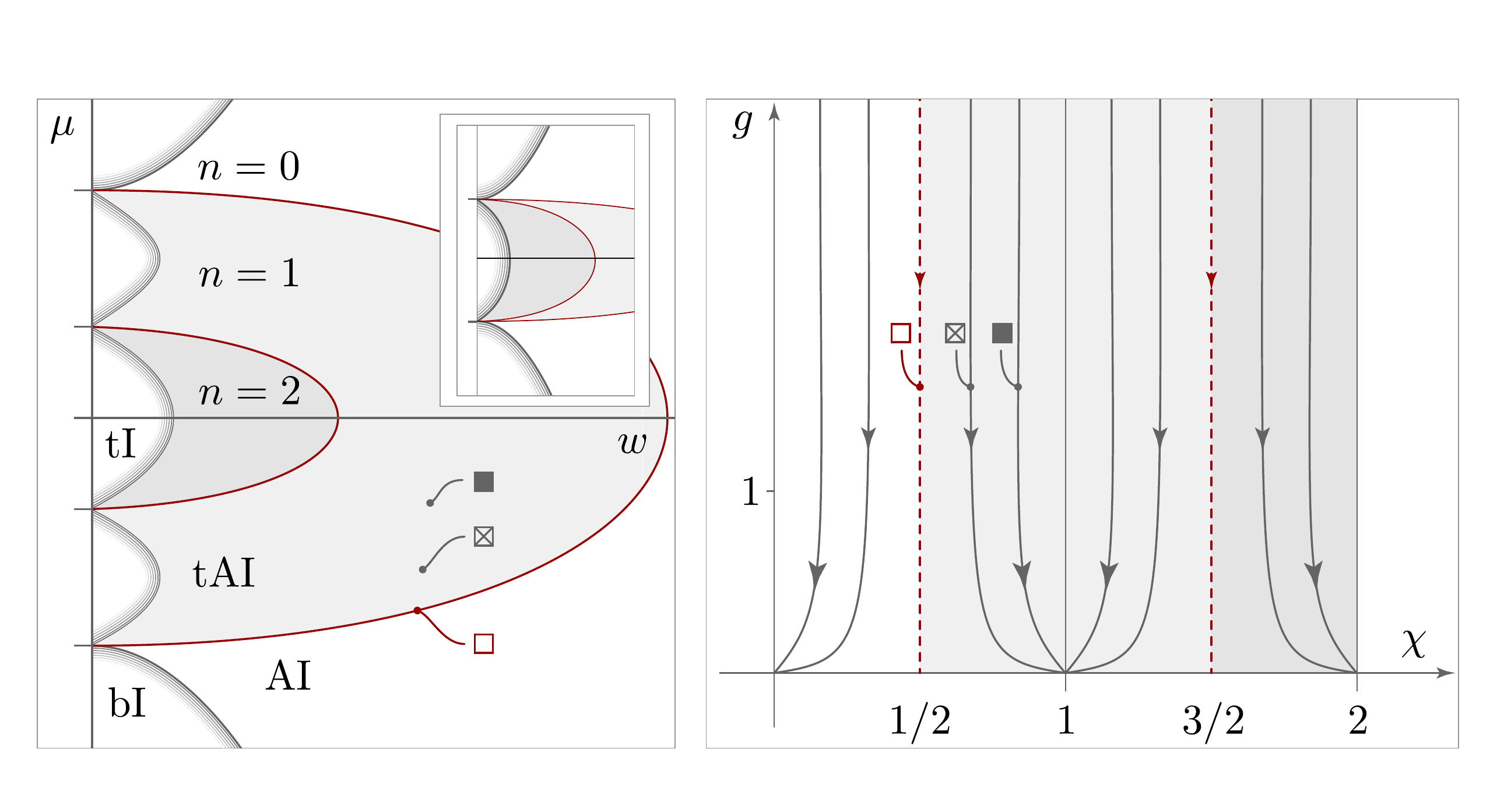}
\caption{Left: $(\mu,w)$ phase diagram of TI. Hatched
areas denote crossover regions between  band and Anderson
insulator. Red lines mark phase transitions between sectors of different
topological index $n$.  The inset describes a situation where the clean system, $w=0$,
has degeneracies, and phase transition points coalesce (e.g. the case of
$N>1$ uncoupled topological chains where inter-chain hopping is solely due to
disorder).  Right: the corresponding $(g,\chi)$-phase diagram. For bare
parameters corresponding to the bulk of a topological phase (cf. the lines
marked by a solid and a crossed box), there is exponentially fast flow in the
system size $L$ towards an insulating configuration: $g=0$ and an integer topological index 
$\chi=n$. At criticality, $\chi+1/2$ (open box), the flow
towards $g=0$ is algebraic in $L$, a signature of a critical state. 
}
\label{fig:PhasePortrait}
\end{figure}

Compared to the enormous research activity on clean topological insulators
(tI), the effects of translational symmetry breaking disorder have begun  to
draw attention only recently~\cite{Li:2009nr,Mondragon-Shem:2013bs,Groth:2009fv,Rieder2013,deGottardi2013}.  In low dimensions, $d\le
2$, the addition of disorder to a topological  insulator drives a
crossover into a topological variant of an Anderson insulator (tAI). The
latter has to be topologically  charged for  the phases carrying different
indices in the clean case cannot simply `disappear' even in the presence of
disorder strong enough to fill the band gap. This means that the  phase
transition points emerging when a control parameter $\mu$ characterizing the
clean system is varied must turn into lines of phase transitions meandering
through a phase plane spanned by $\mu$ and the disorder strength $w$ (cf.
Fig.~\ref{fig:PhasePortrait}.) For a number of tIs, the ensuing phase diagrams
have  been portrayed by numerical methods~\cite{Li:2009nr,Mondragon-Shem:2013bs,Groth:2009fv}, and for one-dimensional  topological
superconductors phase transition points have been identified by transfer
matrix techniques~\cite{Rieder2013,deGottardi2013}.

However, the best studied example of a tAI  remains the  quantum Hall (QH)
insulator. Within the QH context, the crucial role played by
disorder with regard to criticality, edge state formation, and other
phenomena has been appreciated right after the discovery of the
effect~\cite{Prange1981}. Its influence  on the universal physics of the
QH effect is described by Pruisken's theory~\cite{Pruisken1984a}, a
field theory governed by two parameters $(g,\chi)$, where $g$ is a measure of
longitudinal electric conduction, and $\chi$ a topological parameter
proportional to the Hall conductance. At the bare level, both $g$ and $\chi$
are non-universal functions of system parameters, where $g\gg 1$ signifies a
`weakly disordered' system, and half integer values $\chi=n+1/2$ define the
demarcation lines between sectors of different topological index. Increasing
the system size, the parameters $(g,\chi)$ undergo renormalization group flow either
towards tAI states $(0,n)$ with vanishing conductance and
integer Hall angle, or towards  QH transition points
$(g^\ast,n+1/2)$ at criticality. Unfortunately, the fixed points are
buried deeply in the realm of strong coupling, $g^\ast =
\mathcal{O}(1)$, and so far evade analytical treatment.

The  statement made by the present paper is that a nearly identical scenario
repeats itself in quasi-1d disordered topological insulators, viz. the
$\Bbb{Z}$-insulators of symmetry class $\mathrm{AIII}$ and  $\mathrm{BDI}$.
The addition of disorder to quasi-1d insulators of \emph{finite} length $L$
creates intra-gap states, which  turn the system into a conductor,
$g(L)\not=0$, thus compromising its topological integrity~\footnote{The
topological identity is compromised in the sense that the {\em integer} valued
index becomes a non-universal function of the disorder configuration with non-integer 
mean value. Only in the limit of infinite system size does the index
function  self-average to an integer.}. Upon increasing the system size
localization effects kick in, and this leads to the restoration of a self-averaging 
topological index which now is \emph{stabilized} by the presence of
disorder. We describe this reentrance mechanism in terms of a system size
dependent flow of an index function $\chi$, playing a role analogous to the
$2d$ Hall conductance. Our main results are flow diagrams for the two
parameters $(g(L),\chi(L))$ strikingly similar to those of the QH system, but
unlike these fully tractable by analytic means. We will compare to numerical
transfer matrix methods to demonstrate that  the flow diagrams accurately
describe disorder induced quantum criticality.

Before turning to model specific calculations, let us discuss  the structure
of the two-parameter flow in qualitative terms. In the absence of disorder,
our `index' $\chi=n$ reduces to the standard $\Bbb{Z}$-valued winding
numbers~\cite{Schnyder2008} characterizing the clean system. Disorder strong
enough to fill the gap, but too weak to localize, $L<\tilde \xi$, where
$\tilde \xi = N l$ is the localization length, $l\sim |w|^2$  the elastic
scattering mean free path, and $N$ the number of quantum channels carried by
the system, renders the system effectively metallic with Ohmic conductance
$g(L)=\tilde \xi/L$. In the metallic regime, the index $\chi(L)=\tilde
\chi(\mu,w,\dots)\not= n$ assumes a real value where $(\dots)$ denotes non-universal dependence on microscopic system parameters. Upon increasing $L$ the system enters a localization regime, $g(L)\sim \exp(-L/
\xi(\tilde \chi))$, characterized by an effective  length scale $\xi (\tilde
\chi)\sim |\tilde
\chi-n-1/2|^{-\nu}$ where $\nu$ is a correlation length exponent. Along with
the flow $g(L)\to 0$ towards a tAI configuration, the index $\chi(L)$ flows
from its bare value value $\tilde \chi$ back to the nearest integer
$\chi(L)\stackrel{L\to
\infty}\longrightarrow [\tilde \chi]=n$. The exponentially fast two-parameter
flow $(g(L),\chi(L))\to (0,n)+\mathcal{O}(e^{-L/\xi(\tilde \chi)})$ is the quasi-1d analog of the 2d class $\mathrm{A}$ QH scaling (cf.
Fig.~\ref{fig:PhasePortrait}).  Along with the approach to the insulating
state the system stabilizes $n$ zero energy edge states  at
its end. Transitions between distinct topological sectors are marked by half
integer values $\tilde \chi=n+1/2$. At criticality the parameter
$\chi(L)=\tilde \chi=n+1/2$ remains stationary and algebraic scaling $g(L)\sim
L^{-\alpha}$ of the average conductance signifies the presence of a critical 
delocalized state.

The analogy to  QH  extends to the formal level in  that
the quasi-$1d$ insulators, too, are described by a two parameter field theory. On
the bare level, the theory is determined by the pair $(\tilde \xi,\tilde
\chi)$ describing the system at length scales  $l<L<\tilde\xi$. Technically, these
constants are computed from a  self-consistent Born approximation (SCBA) to
the Green function, and criticality is detected by probing half-integerness of $\tilde \chi$. 
We confirm numerically that this procedure 
accurately describes the phase diagram for given models of disorder. 
The running observables $(g(L),\chi(L))$ are then extracted by
probing the  sensitivity of the theory to twists in real space boundary conditions,  an operation analogous to
Pruisken's `background field' method~\cite{Pruisken1987,Pruisken1987a}. Before turning to the  $\mathrm{BDI}$-insulator we  discuss the somewhat simpler $\mathrm{AIII}$
system.

\emph{$\mathrm{AIII}$--insulator ---} Consider a system of $N$ 
chains of length $L$ described by the Hamiltonian,  ${\hat H = \sum_{l}\left[
C^{\dagger}_{l} ((\mu+t)+(\mu-t)\hat P)C_{l+1} +C_l^\dagger \hat V_l C_{l+1} \right] +
\mathrm{h.c.}}$, where $C_l=\{c_{l,j}\}$ is a vector of fermion creation
operators, $j=1,\ldots N$ is the chain index, $l=1,\ldots L$ labels chain sites,  the 
intra-chain hopping is staggered by the parameter $a=|\mu|-t$ and $\hat Pc_{l,j} =
(-)^{l} c_{l,j}$ acts as a `parity operator'. The matrix $\hat V_l$ describes the random inter-chain hopping, 
which is Gaussian distributed with correlators 
$\langle V^{ij}_l (V^{i'j'}_{l'})^* \rangle=(w^2/N) \delta_{ij} \delta_{i'j'} \delta_{ll'}$. 
The anti-commutativity of our (time-reversal non-invariant) Hamiltonian with
sublattice parity, $[\hat H,\hat P]_+=0$ indicates that the system belongs to
symmetry class $\mathrm{AIII}$. In the absence of disorder, a topolgical
index, or `winding number' may be defined as~\cite{Schnyder2008} $n\equiv
-i\sum_{q=1}^N\int_0^{2\pi}\frac{dk}{2\pi}\, \mathrm{tr}\,(Q^{-1}\partial_k
Q)$, where $Q\equiv H_{+-}$ is the block of the Hamiltonian connecting sites
of positive and negative parity, and  $k,q$ are Fourier indices,
${c_{k,q}=\frac{1}{LN}\sum_{l,j} e^{i(kl/L+q2 \pi j/N)} c_{l,j}}$. 
Assuming real hopping amplitude $t>0$ for simplicity, a
straightforward calculation shows that $n =N \Theta(t-|\mu|)$, which identifies
the amplitude $\mu$ as a topological control
parameter triggering a transition from $n=0$ to $n=N$ at 
$\mu=\pm t$.~\footnote{In the presence of non-random inter-chain coupling the
transition point would split into $N$ points of unit change in $n$, cf.
Fig.~\ref{fig:PhasePortrait} main panel vs. inset.} To access the index $n$ in
a way not  tied to translational invariance, we imagine the chains
compactified to a ring of radius $L$ and pierced by a staggered flux $\phi_0$
which affects the fermions as $c_{l,j}\to e^{i\hat P \phi_0 l/L}c_{l,j}$ and 
the momentum representation of the chiral blocks as
$Q(k)\to Q(k+\phi_0)$ and $Q^\dagger(k)\to Q^\dagger(k-\phi_0)$, resp. Now
consider the generating function ${Z(
\phi)=\det(\hat G^{-1}_0(\phi_0))/\det(\hat G^{-1}_0(-i\phi_1))}$, where ${
\phi=(\phi_0,\phi_1)^T}$ and $\hat G_\epsilon(\phi_\alpha)=(\epsilon^+-\hat
H(\phi_\alpha))^{-1}$  is the retarded Green function of the gauged  Hamiltonian.
The  transformation of $Q$ then implies  the representation $n\equiv
\chi= \frac{i}{2}\partial_{\phi_0}\big|_{    \phi=(0,0)}Z(    \phi)$, which no longer
relies on the momentum space language. In this formula we also anticipate that in non-translational invariant systems $n$ may generalize to a non-integer
parameter $\chi$.

\emph{Field theory ---} We next ask how the system responds to the presence of
disorder. The generating function  averaged over a Gaussian distribution of the bond amplitudes $V_{j,l}$ affords
a representation in terms of a nonlinear $\sigma$-model, ${Z(    \phi)=\int {\cal D}T \,\exp(-S[T])}$, with the
action~\cite{Altland2001511}
\begin{equation}
\label{eq:AIIIAction}    
S[T]=\!
\int\limits_0^{L} \!\! dx \left[\frac{\tilde \xi}{4}\,\mathrm{str}(\partial_x T\partial_x T^{-1})
+ \tilde\chi\,\mathrm{str}(T^{-1}\partial_x T)\right].  
\end{equation}   
Here, `$\mathrm{str}$' is the supertrace and $T=U\left(\begin{smallmatrix} e^{y_1} & \crcr  & e^{iy_0} \end{smallmatrix}
\right)U^{-1}$ is a $2\times2 $ supermatrix field
parameterizing the field space $\mathrm{GL}(1|1)$~\cite{Zirnbauer1996} in terms of two radial
coordinates $y_1\in \Bbb{R}$, $y_0\in[0,2\pi[$ and two Grassmann valued angular variables
$\rho,\sigma$, where $U=\exp \left(\begin{smallmatrix} & \rho \crcr \sigma &  \end{smallmatrix}
\right)$. The theory contains two coupling constants,
the localization length $\tilde\xi$ and the coefficient of the
topological term $\tilde\chi$. The latter is computed from the underlying microscopic theory as~\cite{Altland2001511} $ \tilde \chi= \frac{i}{2}\mathrm{tr}(\hat G^+P \partial_k \hat H)$,
where $\hat G^+$ now stands for the zero energy Green function averaged over
disorder within  self consistent Born approximation. For vanishingly weak disorder, $G^+\to
-\hat H^{-1}$
and $\tilde \chi\to n$ reduces to the winding number. The value of $\tilde
\chi$ in the presence of disorder depends on model specifics and will be discussed in more concrete terms below.  The external parameters
$\phi$ enter the theory through a twisted boundary condition, $T(L)=
\mathrm{diag}(e^{\phi_1},e^{i\phi_0})$, and $T(0)=\mathds{1}$. 
We finally note that the boundary shift in the non-compact `angle', $\phi_1$
enables us to extract the conductance of the wire
as~\cite{Nazarov1994,Lamacraft2004} $g=(\partial^2_0+\partial^2_1)|_{
\phi=(0,0)} Z$, where $\partial_\alpha=\partial_{\phi_\alpha}$ for brevity.
The rationale behind this expression is that the second derivatives
$\partial^2_\alpha Z\sim
\partial^2_\phi \ln\det(G_0(\phi_\alpha))$ probe the average `curvature' of the
$    \phi$-dependent energy levels which, according to Thouless~\cite{Thouless1974}, is a measure of
the system's conductance.

\begin{figure}
\centering{
\includegraphics[width=0.4\textwidth]{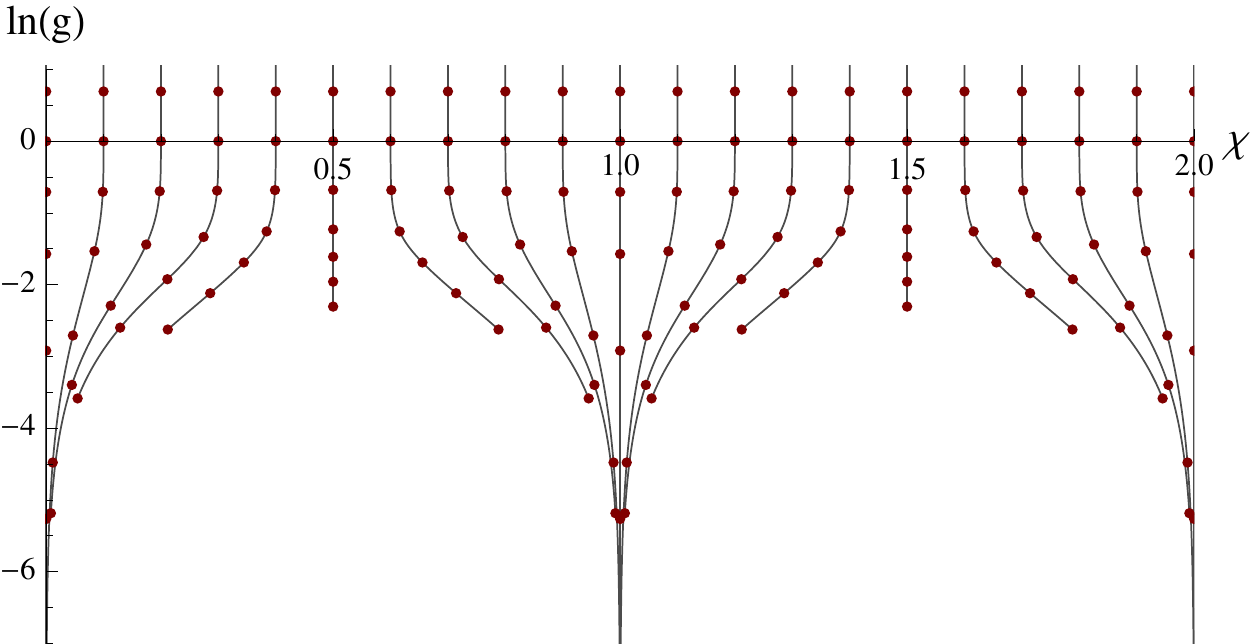}
}
\caption{Flow of the conductance $g$ and the topological parameter $\chi$ as a
function of system size. Dots are for values,
$L/    \tilde\xi=\frac{1}{4},\frac{1}{2},1,2\dots,32$.}
\label{fig:FlowGraph}
\end{figure}

Our goal is to understand the scaling of the observables $(g,\chi)$ in dependence
on the system size $L$. In the metallic regime $    l<L\ll
\tilde\xi$, `size quantization' of the operator $\partial_x T \partial_x
T^{-1}$ suppresses fluctuations of $T$. Substitution of the minimal
configuration compatible with the boundary conditions,
$T(x)=\mathrm{diag}(e^{\phi_1x/L},e^{i\phi_0 x/L})$ then yields $(g,\chi)=(
\tilde\xi/L,\tilde \chi)$, i.e. an Ohmic conductance, and a topological index
set by the bare value $\tilde \chi$. To understand what happens upon entering
the localization regime, $L\gtrsim     \tilde\xi$, it is convenient to think
of $x$ as imaginary time, and of $Z(    \phi)$ as the path integral
describing the free motion (first term in the action
Eq.~\eqref{eq:AIIIAction}) of a fictitious quantum particle on the manifold
$\mathrm{GL}(1|1)$ in the presence of a constant gauge flux (the second term).
The
time-dependent ``Schr\"odinger''
equation corresponding to the path integral~\cite{Altland2001511}
\begin{equation}
                                                               \label{eq:Schrodinger}
    \tilde\xi\, \partial_x \Psi(     y,x)=\frac{1}{J(y)}(\partial_\alpha-iA_\alpha)J(     y)(\partial_\alpha-iA_\alpha) \Psi(     y,x),
\end{equation}   
is governed by the  Laplacian, $J^{-1}\partial_\alpha J \partial_\alpha$
on  $\mathrm{GL}(1|1)$, where ${J(     y)=\sinh^{-2}\left(\frac{1}{2}( y_1-i
y_0)\right)}$ is a Jacobian  and the coupling to the vector potential, $
A=\tilde \chi(1, i)^T$ represents the topological term. From
Eq.~\eqref{eq:Schrodinger} we obtain the partition function as
$Z(\phi)=\Psi(\phi,L)$. The equation can be solved by spectral decomposition
${\Psi(\phi,L)=1 + \sum_{l_0\in \Bbb{Z}+1/2} \int \frac{dl_1}{(2\pi)^2}\,
P_{l}\, \Psi_{l}(    \phi)\, e^{-\epsilon_{l}L/\tilde\xi}}$, where `1' is by
supersymmetric normalization of the partition function $Z(0,x)=1$, and
$P_l=4\pi/(l_0+il_1)$ implements the the spectral decomposition of the initial
condition $\Psi(\phi,x\to 0)$. The  eigenfunctions of
Eq.~(\ref{eq:Schrodinger}) are given by $\Psi_{l}(     y) =\sinh\left
(\frac{1}{2} ( y_1-i y_0)\right)e^{il_\alpha y_\alpha} $, and the
corresponding eigenvalues by
$
\epsilon_{l}=  (l_0-\tilde\chi)^2+(l_1-i\tilde \chi)^2.                                                           
$
From this representation, it is straightforward to compute the first and second
order expansions in $\phi$ to arrive at the result
\begin{eqnarray}
 g&=&\sqrt{\frac{\tilde\xi}{\pi L}}\sum_{l_0\in \Bbb{Z}+1/2} e^{-(l_0-\tilde\chi)^2 L/\tilde\xi},\\
  \chi&=& n-\frac{1}{4}\!\sum_{l_0\in \Bbb{Z}+1/2}\! \left[\mathrm{erf}\Big(\sqrt{\frac{L}{\tilde\xi}}\,(l_0-\delta\tilde\chi)\Big)-(\delta\tilde\chi \leftrightarrow
  -\delta\tilde\chi)\right],\nonumber
 \end{eqnarray}
where $\delta\tilde\chi=\tilde\chi-n$ is the deviation of $\tilde\chi$ off the nearest
integer value, $n$. For generic bare values $(    \tilde\xi, \tilde \chi)$
the two formulas describe an exponentially fast approach towards an insulating
state $(0,n)$ upon increasing length $L$. At criticality, $(\tilde\xi,n+1/2)$, the
topological angle remains invariant, while an algebraic decay of the
conductance $g(L)\approx (\tilde\xi/\pi L)^{1/2}$ signifies the presence
of a delocalized state at the band center. The emergence of power law scaling
at criticality can be described in terms of an effective correlation length $
\xi(\chi)=    \tilde\xi |\chi-n-1/2|^{-\nu}$. Comparing the ansatz, $g\sim
\exp(-L/\xi(\chi))$, to the result above, we identify the correlation
length exponent $\nu=2$~\footnote{Notice that the exponent describing the  transport coefficient $\protect{\sim \langle \exp(-L/\xi)\rangle}$ differs from the exponent $\protect{\nu=1}$ for the average correlation 
length, $\protect{\left\langle \xi \right\rangle}$.\cite{BMSA:98,Mondragon-Shem:2013bs}}. A number of flow lines are shown graphically in
Fig.~\ref{fig:FlowGraph}. which is the 1d analogue of the two-parameter flow
diagram~\cite{Khmelnitskii83} describing criticality in the integer QH system.

\emph{Class $\mathrm{BDI}$ ---} We next extend our discussion
to the presence of time reversal, symmetry class $\mathrm{BDI}$. System of this type are realized, e.g.~\cite{Altland1997}, as lattice $p$-wave superconductors with Hamiltonian
$\hat H = \sum_{l=1}^L [C^\dagger_l \hat H_{0,l} C_l+(C^\dagger_l \hat
H_{1,l} C_{l+1}+\mathrm{h.c.})]$, where the spinless fermion operators
$C_l=(c_{l,j},c_{l,j}^\dagger)^T$ are vectors in channel and Nambu spaces. The
on-site part of the Hamiltonian,  ${\hat H_{0,l} =
(\mu/2 + \hat V_{l}) \sigma_3}$ contains the chemical
potential $\mu$ and real symmetric inter-chain matrices  $\hat
V_l$; 
$\sigma_i$ acting in Nambu space. The contribution, $\hat
H_{1,l}=-\frac{1}{2}t_l
\sigma_3+\frac{i}{2}\hat \Delta_l \sigma_2$, contains nearest neighbor hopping, $t_l$,  and 
the order parameter, $\hat \Delta_l$, here assumed to be imaginary for
convenience. Quantities carrying a subscript
`$l$' may contain site-dependent random contributions. The first quantized representation of $\hat H$
obeys the chiral symmetry $[\hat P,\hat H]_+=0$, with $\hat P=\sigma_1$. 
The clean system  supports $n\le N$ Majorana end
states, where $n$ decreases  upon increasing $\mu$.
Generalizing to the presence of disorder, we obtain a pattern of phase
transition lines similar to the one discussed above. Before turning to field
theory, we apply transfer matrix methods to a numerical description of the
ensuing phase portrait.

\begin{figure}
\centering{
\includegraphics[width=0.32\textwidth]{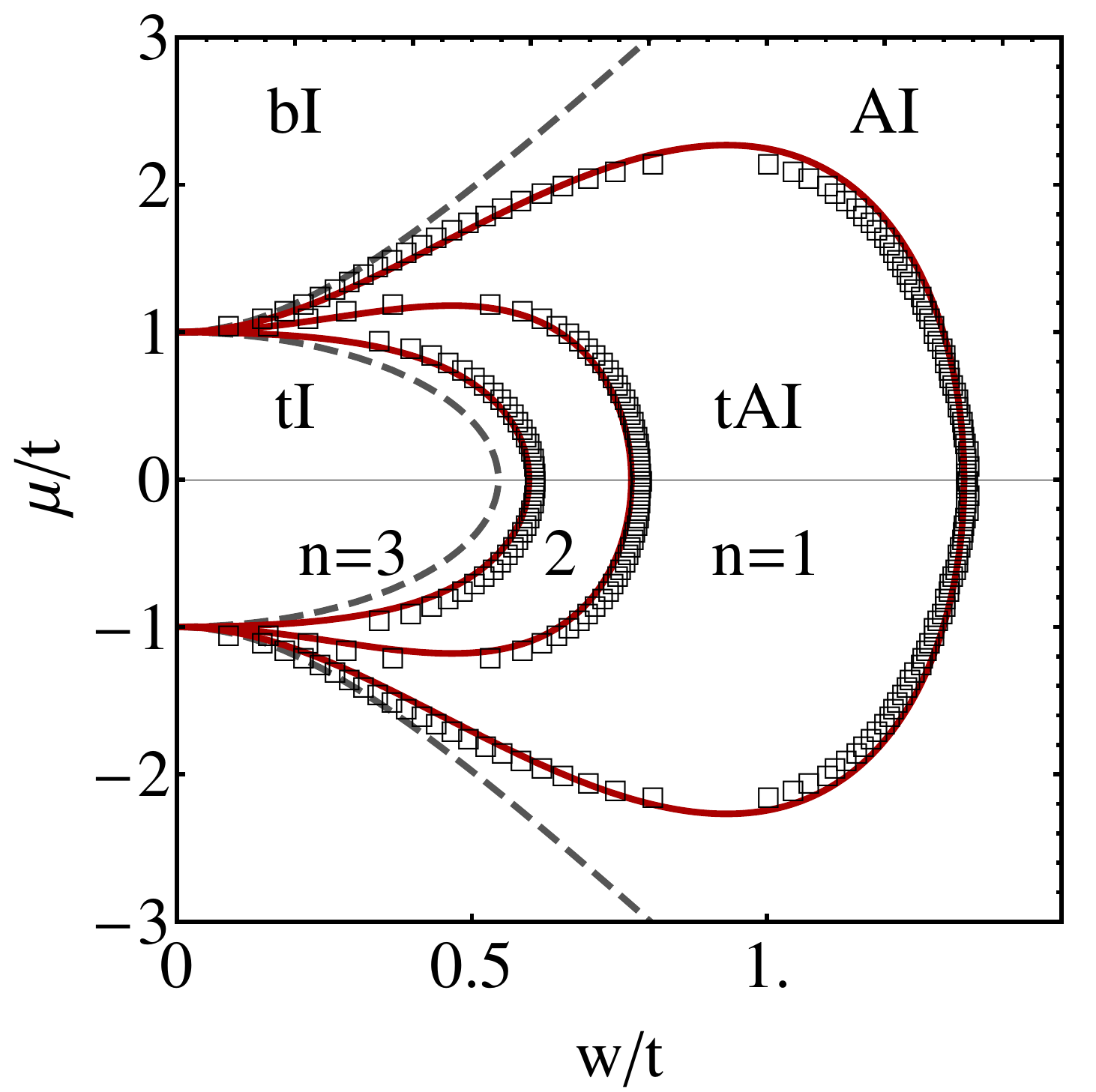}
}
\caption{Phase diagram of the class BDI 3-channel wire.
Dashed lines show crossover regions between bI and AI or tI and tAI phases, 
derived from the SCBA. Solid lines correspond to half integer values of the SCBA computed index $\tilde\chi$ and mark boundaries between phases of different $n$. 
BI and AI have $n=0$, while $n>0$ for tI and tAI. Data points are phase boundaries found 
from a numerical analysis of Lyapunov exponents $\lambda_j$.}
\label{fig:BDI_D_N}
\end{figure}

\emph{Transfer Matrix ---} Defining a doublet $\eta_l=(\psi_{l+1},\psi_l)^T$, one  verifies that the zero energy eigenfunctions, $\psi_l$ of the lattice Hamiltonian  obey
the recursion relation
$    \eta_{l+1}=\mathcal{T}_l     \eta_l$, where
$
 \mathcal{T}_l=\left ( \begin{smallmatrix} -H_{1,l}^{-1}H_{0,l}  & - H_{1,l}^{-1}  H_{1,l}^\dagger\\ 1 & 0 \end{smallmatrix} \right),
 $
 and we assumed non-degeneracy of the hopping matrices $\{H_{1,l}\}$. We
iterate this equation to obtain $    \eta_L=\mathcal{T}    \eta_1$,
where $\mathcal{T}=\prod_{l}^L\mathcal{T}_l$ is the transfer matrix. The presence of a chiral structure
means that $\mathcal{T}=\mathrm{bdiag}(\mathcal{T}^{11},\mathcal{T}^{22})$ can be brought to a block-diagonal form. Representing 
the positive eigenvalues of $\mathcal{T}^{11}$ as $\exp(L \lambda_j)$, $j=1,\dots,N$,  the
index $n$ of individual systems is given by the the number of negative
`Lyapunov exponents', $\lambda_j<0$~\cite{Fulga2011}. We numerically compute the average of these numbers by sampling from a  Gaussian distribution of on-site potentials, 
$\hat V_l = \hat V_{0,l} \sigma_0 + \hat V_{3,l} \sigma_3$, with correlators 
$\langle V_{a,l}^{ij} V_{a,l}^{i'j'} \rangle = (w^2/N) ( \delta_{ii'} \delta_{jj'} - (-)^a \delta_{ij'} \delta_{ji'})$, $a=0,3$ 
and for a model with $t=\Delta$. The results are shown in Fig.~\ref{fig:BDI_D_N}, where boxes indicate 
changes of the topological number. 

\emph{Field Theory ---} The field theoretical description of the system 
parallels that of the $\mathrm{AIII}$-insulator. Time reversal invariance means
that the fields are now $4\times 4$ matrices spanning the coset space
$\mathrm{GL}(2|2)/\mathrm{OSp}(2|2)$~\cite{Zirnbauer1996}, the field action is
given by Eq.~\eqref{eq:AIIIAction} as before. 
As in class $\mathrm{AIII}$, the topological coupling constant is given by $\tilde \chi=\frac{i}{2}\mathrm{tr}(\hat
G^+\hat{P}\partial_k
\hat H)$ where the 
Green function $\hat G^+=(i0 -\hat H -\hat \Sigma)^{-1}$ contains a self
energy $\hat \Sigma = -i \Sigma_0 \sigma_0 + \Sigma_3\sigma_3$. For random 
 $\hat V_l$ (as in the transfer matrix study) the latter is determined by the
 SCBA equation
$-i \hat \Sigma_a = i^k w^2 {\rm tr}( \hat G^+_{lj,lj} \sigma_a)$, where the 
trace is over the Nambu space. 
This algebraic equation can be
solved numerically, and as a result we obtain  contour lines of half integer
$\tilde \chi$ providing an excellent approximation to the numerical transfer matrix data (cf. Fig.~\ref{fig:BDI_D_N}). 
In select regions of interest the SCBA equations can be solved analytically. For
example, we find that the phase transition points on the $(\mu=0,w)$ abscissa
are located at $t(N/(2n+1))^{1/2}$~\cite{Rieder2013}, where $0\leq n<N$, or that for disorder weaker than these values the degenerate
phase transition point $(\mu,w)=(\Delta,0)$ on the clean system ordinate fans out into  $N$ lines $(\mu_n,w)=\big(\Delta+\frac{2w^2}{\Delta}(3-\frac{4n+2}{N}),w\big)$.

The observable pair $(g,\chi)$ can be extracted from the field theory as in
the $\mathrm{AIII}$-theory. The manifold of $T$-matrices is now spanned by $8$
coordinates, three of which, $(y_0,y_1,y_2)$, $y_0\in[0,2\pi[$, $y_1\in
\Bbb{R}, y_2\in \Bbb{R}^+$ play a role analogous to the radial coordinates of
the $\mathrm{AIII}$--manifold. Eq.~\eqref{eq:Schrodinger}  has to be solved at
$y=(\phi_0,\phi_1,0)$, the Jacobian is given by $J(y)=\sinh(2 y_2)/(16
\sinh^2(y_1-iy_0+y_2)\sinh^2(y_1-iy_0-y_2))$, and the vector potential by
$A=2\tilde \chi(1,i,0)^T$. Although we are not able to solve this equation in
generality, we observe that at large values of the variables $y_{1,2}$ the
$\sinh$-functions simplify to exponentials. The wave functions in this regime,
too, show an exponential profile from which the eigenvalues follow as
$\epsilon_l=\left[(l_0 - 2\tilde \chi)^2+(l_1-2i\tilde
\chi)^2+l_2^2+1\right]/4$, with integer $l_0$ and real $l_{1,2}$. While for
general $\tilde \chi$ the positive real part of $\epsilon_l$ signals
localization, we have  a zero eigenvalue $\epsilon_{(1,0,0)}=0$ at  $\tilde
\chi=1/2$. This signals delocalization, and criticality. The $(g,\chi)$ flow diagram therefore has the same 
structure as in Fig.~\ref{fig:PhasePortrait}b.

\emph{Discussion ---} We have shown that quantum criticality in the quasi 1d $\Bbb{Z}$-insulators of class $\mathrm{AIII}$, and $\mathrm{BDI}$ is governed by a two parameter flow diagram strikingly similar to that of 2d quantum Hall effect. The flow describes the system-size dependence
of two parameters -- the average conductance, and a
topological signature -- whose initial values  are  functions of the system's microscopic
parameters. This correspondence enables us to  describe
criticality in  disordered topological systems in quantitative agreement with numerical simulations.  The two parameter field theories behind this
picture describe the universal aspects of  both the surface physics and localization in the bulk at
length scales exceeding the mean free path. We conjecture that the same scenario of two parameter field theory/flow diagram holds for other tAI: 2d superconducting classes C and D and possibly 3d CI, DIII and AIII Anderson insulators.        Exploring to which extent the
current framework extends to the $\Bbb{Z}_2$ topological insulators,  requires
further research.

{\it Acknowledgment} Discussions with P. Brouwer are gratefully acknowledged.
Work supported by SFB/TR 12 of the Deutsche Forschungsgemeinschaft. AK was supported by NSF grant DMR1306734. LF acknowledges support from the D-ITP consortium, a program of the Netherlands Organization for Scientific Research (NWO) that is funded by the Dutch Ministry of Education, Culture and Science (OCW) as well as from the Deutsche Forschungsgemeinschaft under FR 2627/3-1.


\begin{thebibliography}{22}%
\makeatletter
\providecommand \@ifxundefined [1]{%
 \@ifx{#1\undefined}
}%
\providecommand \@ifnum [1]{%
 \ifnum #1\expandafter \@firstoftwo
 \else \expandafter \@secondoftwo
 \fi
}%
\providecommand \@ifx [1]{%
 \ifx #1\expandafter \@firstoftwo
 \else \expandafter \@secondoftwo
 \fi
}%
\providecommand \natexlab [1]{#1}%
\providecommand \enquote  [1]{``#1''}%
\providecommand \bibnamefont  [1]{#1}%
\providecommand \bibfnamefont [1]{#1}%
\providecommand \citenamefont [1]{#1}%
\providecommand \href@noop [0]{\@secondoftwo}%
\providecommand \href [0]{\begingroup \@sanitize@url \@href}%
\providecommand \@href[1]{\@@startlink{#1}\@@href}%
\providecommand \@@href[1]{\endgroup#1\@@endlink}%
\providecommand \@sanitize@url [0]{\catcode `\\12\catcode `\$12\catcode
  `\&12\catcode `\#12\catcode `\^12\catcode `\_12\catcode `\%12\relax}%
\providecommand \@@startlink[1]{}%
\providecommand \@@endlink[0]{}%
\providecommand \url  [0]{\begingroup\@sanitize@url \@url }%
\providecommand \@url [1]{\endgroup\@href {#1}{\urlprefix }}%
\providecommand \urlprefix  [0]{URL }%
\providecommand \Eprint [0]{\href }%
\providecommand \doibase [0]{http://dx.doi.org/}%
\providecommand \selectlanguage [0]{\@gobble}%
\providecommand \bibinfo  [0]{\@secondoftwo}%
\providecommand \bibfield  [0]{\@secondoftwo}%
\providecommand \translation [1]{[#1]}%
\providecommand \BibitemOpen [0]{}%
\providecommand \bibitemStop [0]{}%
\providecommand \bibitemNoStop [0]{.\EOS\space}%
\providecommand \EOS [0]{\spacefactor3000\relax}%
\providecommand \BibitemShut  [1]{\csname bibitem#1\endcsname}%
\let\auto@bib@innerbib\@empty
\bibitem [{\citenamefont {Li}\ \emph {et~al.}(2009)\citenamefont {Li},
  \citenamefont {Chu}, \citenamefont {Jain},\ and\ \citenamefont
  {Shen}}]{Li:2009nr}%
  \BibitemOpen
  \bibfield  {author} {\bibinfo {author} {\bibfnamefont {J.}~\bibnamefont
  {Li}}, \bibinfo {author} {\bibfnamefont {R.-L.}\ \bibnamefont {Chu}},
  \bibinfo {author} {\bibfnamefont {J.~K.}\ \bibnamefont {Jain}}, \ and\
  \bibinfo {author} {\bibfnamefont {S.-Q.}\ \bibnamefont {Shen}},\ }\href
  {\doibase 10.1103/PhysRevLett.102.136806} {\bibfield  {journal} {\bibinfo
  {journal} {Phys. Rev. Lett.}\ }\textbf {\bibinfo {volume} {102}},\ \bibinfo
  {pages} {136806} (\bibinfo {year} {2009})}\BibitemShut {NoStop}%
\bibitem [{\citenamefont {Mondragon-Shem}\ \emph {et~al.}(2013)\citenamefont
  {Mondragon-Shem}, \citenamefont {Song}, \citenamefont {Hughes},\ and\
  \citenamefont {Prodan}}]{Mondragon-Shem:2013bs}%
  \BibitemOpen
  \bibfield  {author} {\bibinfo {author} {\bibfnamefont {I.}~\bibnamefont
  {Mondragon-Shem}}, \bibinfo {author} {\bibfnamefont {J.}~\bibnamefont
  {Song}}, \bibinfo {author} {\bibfnamefont {T.~L.}\ \bibnamefont {Hughes}}, \
  and\ \bibinfo {author} {\bibfnamefont {E.}~\bibnamefont {Prodan}},\ }\href
  {http://arxiv.org/abs/1311.5233} 
  (\bibinfo {year} {2013}),\ \Eprint {http://arxiv.org/abs/1311.5233}
  {arXiv:1311.5233} \BibitemShut {NoStop}%
\bibitem [{\citenamefont {Groth}\ \emph {et~al.}(2009)\citenamefont {Groth},
  \citenamefont {Wimmer}, \citenamefont {Akhmerov}, \citenamefont
  {Tworzydło},\ and\ \citenamefont {Beenakker}}]{Groth:2009fv}%
  \BibitemOpen
  \bibfield  {author} {\bibinfo {author} {\bibfnamefont {C.}~\bibnamefont
  {Groth}}, \bibinfo {author} {\bibfnamefont {M.}~\bibnamefont {Wimmer}},
  \bibinfo {author} {\bibfnamefont {A.}~\bibnamefont {Akhmerov}}, \bibinfo
  {author} {\bibfnamefont {J.}~\bibnamefont {Tworzydło}}, \ and\ \bibinfo
  {author} {\bibfnamefont {C.}~\bibnamefont {Beenakker}},\ }\href {\doibase
  10.1103/PhysRevLett.103.196805} {\bibfield  {journal} {\bibinfo  {journal}
  {Phys. Rev. Lett.}\ }\textbf {\bibinfo {volume} {103}},\ \bibinfo
  {pages} {196805} (\bibinfo {year} {2009})}\BibitemShut {NoStop}%
\bibitem [{\citenamefont {Rieder}\ \emph {et~al.}(2013)\citenamefont {Rieder},
  \citenamefont {Brouwer},\ and\ \citenamefont {Adagideli}}]{Rieder2013}%
  \BibitemOpen
  \bibfield  {author} {\bibinfo {author} {\bibfnamefont {M.-T.}\ \bibnamefont
  {Rieder}}, \bibinfo {author} {\bibfnamefont {P.~W.}\ \bibnamefont {Brouwer}},
  \ and\ \bibinfo {author} {\bibfnamefont {I.}~\bibnamefont {Adagideli}},\
  }\href {\doibase 10.1103/PhysRevB.88.060509} {\bibfield  {journal} {\bibinfo
  {journal} {Phys. Rev. B}\ }\textbf {\bibinfo {volume} {88}},\ \bibinfo
  {pages} {060509} (\bibinfo {year} {2013})}\BibitemShut {NoStop}%
\bibitem [{\citenamefont {DeGottardi}\ \emph {et~al.}(2013)\citenamefont
  {DeGottardi}, \citenamefont {Sen},\ and\ \citenamefont
  {Vishveshwara}}]{deGottardi2013}%
  \BibitemOpen
  \bibfield  {author} {\bibinfo {author} {\bibfnamefont {W.}~\bibnamefont
  {DeGottardi}}, \bibinfo {author} {\bibfnamefont {D.}~\bibnamefont {Sen}}, \
  and\ \bibinfo {author} {\bibfnamefont {S.}~\bibnamefont {Vishveshwara}},\
  }\href {\doibase 10.1103/PhysRevLett.110.146404} {\bibfield  {journal}
  {\bibinfo  {journal} {Phys. Rev. Lett.}\ }\textbf {\bibinfo {volume}
  {110}},\ \bibinfo {pages} {146404} (\bibinfo {year} {2013})}\BibitemShut
  {NoStop}%
\bibitem [{\citenamefont {Prange}(1981)}]{Prange1981}%
  \BibitemOpen
  \bibfield  {author} {\bibinfo {author} {\bibfnamefont {R.}~\bibnamefont
  {Prange}},\ }\href {\doibase 10.1103/PhysRevB.23.4802} {\bibfield  {journal}
  {\bibinfo  {journal} {Phys. Rev. B}\ }\textbf {\bibinfo {volume} {23}},\
  \bibinfo {pages} {4802} (\bibinfo {year} {1981})}\BibitemShut {NoStop}%
\bibitem [{\citenamefont {Pruisken}(1984)}]{Pruisken1984a}%
  \BibitemOpen
  \bibfield  {author} {\bibinfo {author} {\bibfnamefont {A.}~\bibnamefont
  {Pruisken}},\ }\href
  {http://www.sciencedirect.com/science/article/pii/0550321384901019}
  {\bibfield  {journal} {\bibinfo  {journal} {N. Phys. B}\ }\textbf
  {\bibinfo {volume} {235}},\ \bibinfo {pages} {277} (\bibinfo {year}
  {1984})}\BibitemShut {NoStop}%
\bibitem [{Note1()}]{Note1}%
  \BibitemOpen
  \bibinfo {note} {The topological identity is compromised in the sense that
  the {\protect \em integer} valued index becomes a non-universal function of
  the disorder configuration with non-integer mean value. Only in the limit of
  infinite system size does the index function self-average to an
  integer.}\BibitemShut {Stop}%
\bibitem [{\citenamefont {Schnyder}\ \emph {et~al.}(2008)\citenamefont
  {Schnyder}, \citenamefont {Ryu}, \citenamefont {Furusaki},\ and\
  \citenamefont {Ludwig}}]{Schnyder2008}%
  \BibitemOpen
  \bibfield  {author} {\bibinfo {author} {\bibfnamefont {A.}~\bibnamefont
  {Schnyder}}, \bibinfo {author} {\bibfnamefont {S.}~\bibnamefont {Ryu}},
  \bibinfo {author} {\bibfnamefont {A.}~\bibnamefont {Furusaki}}, \ and\
  \bibinfo {author} {\bibfnamefont {A.}~\bibnamefont {Ludwig}},\ }\href
  {\doibase 10.1103/PhysRevB.78.195125} {\bibfield  {journal} {\bibinfo
  {journal} {Phys. Rev. B}\ }\textbf {\bibinfo {volume} {78}},\ \bibinfo
  {pages} {195125} (\bibinfo {year} {2008})}\BibitemShut {NoStop}%
\bibitem [{\citenamefont {Pruisken}(1987{\natexlab{a}})}]{Pruisken1987}%
  \BibitemOpen
  \bibfield  {author} {\bibinfo {author} {\bibfnamefont {A.~M.}\ \bibnamefont
  {Pruisken}},\ }\href {\doibase 10.1016/0550-3213(87)90363-4} {\bibfield
  {journal} {\bibinfo  {journal} {N. Phys. B}\ }\textbf {\bibinfo
  {volume} {285}},\ \bibinfo {pages} {719} (\bibinfo {year}
  {1987}{\natexlab{a}})}\BibitemShut {NoStop}%
\bibitem [{\citenamefont {Pruisken}(1987{\natexlab{b}})}]{Pruisken1987a}%
  \BibitemOpen
  \bibfield  {author} {\bibinfo {author} {\bibfnamefont {A.~M.}\ \bibnamefont
  {Pruisken}},\ }\href {\doibase 10.1016/0550-3213(87)90178-7} {\bibfield
  {journal} {\bibinfo  {journal} {N. Phys. B}\ }\textbf {\bibinfo
  {volume} {290}},\ \bibinfo {pages} {61} (\bibinfo {year}
  {1987}{\natexlab{b}})}\BibitemShut {NoStop}%
\bibitem [{Note2()}]{Note2}%
  \BibitemOpen
  \bibinfo {note} {In the presence of non-random inter-chain coupling the
  transition point would split into $N$ points of unit change in $n$, cf.
  Fig.~\ref {fig:PhasePortrait} main panel vs. inset.}\BibitemShut {Stop}%
\bibitem [{\citenamefont {Altland}\ and\ \citenamefont
  {Merkt}(2001)}]{Altland2001511}%
  \BibitemOpen
  \bibfield  {author} {\bibinfo {author} {\bibfnamefont {A.}~\bibnamefont
  {Altland}}\ and\ \bibinfo {author} {\bibfnamefont {R.}~\bibnamefont
  {Merkt}},\ }\href {\doibase http://dx.doi.org/10.1016/S0550-3213(01)00209-7}
  {\bibfield  {journal} {\bibinfo  {journal} {N. Phys. B}\ }\textbf
  {\bibinfo {volume} {607}},\ \bibinfo {pages} {511} (\bibinfo {year}
  {2001})}\BibitemShut {NoStop}%
\bibitem [{\citenamefont {Zirnbauer}(1996)}]{Zirnbauer1996}%
  \BibitemOpen
  \bibfield  {author} {\bibinfo {author} {\bibfnamefont {M.~R.}\ \bibnamefont
  {Zirnbauer}},\ }\href {\doibase 10.1063/1.531675} {\bibfield  {journal}
  {\bibinfo  {journal} {J. of Math. Phys.}\ }\textbf {\bibinfo
  {volume} {37}},\ \bibinfo {pages} {4986} (\bibinfo {year}
  {1996})}\BibitemShut {NoStop}%
\bibitem [{\citenamefont {Nazarov}(1994)}]{Nazarov1994}%
  \BibitemOpen
  \bibfield  {author} {\bibinfo {author} {\bibfnamefont {Y.}~\bibnamefont
  {Nazarov}},\ }\href {\doibase 10.1103/PhysRevLett.73.134} {\bibfield
  {journal} {\bibinfo  {journal} {Phys. Rev. Lett.}\ }\textbf {\bibinfo
  {volume} {73}},\ \bibinfo {pages} {134} (\bibinfo {year} {1994})}\BibitemShut
  {NoStop}%
\bibitem [{\citenamefont {Lamacraft}\ \emph {et~al.}(2004)\citenamefont
  {Lamacraft}, \citenamefont {Simons},\ and\ \citenamefont
  {Zirnbauer}}]{Lamacraft2004}%
  \BibitemOpen
  \bibfield  {author} {\bibinfo {author} {\bibfnamefont {A.}~\bibnamefont
  {Lamacraft}}, \bibinfo {author} {\bibfnamefont {B.}~\bibnamefont {Simons}}, \
  and\ \bibinfo {author} {\bibfnamefont {M.}~\bibnamefont {Zirnbauer}},\ }\href
  {\doibase 10.1103/PhysRevB.70.075412} {\bibfield  {journal} {\bibinfo
  {journal} {Phys. Rev. B}\ }\textbf {\bibinfo {volume} {70}},\ \bibinfo
  {pages} {075412} (\bibinfo {year} {2004})}  \BibitemShut
  {NoStop}%
\bibitem [{\citenamefont {Thouless}(1974)}]{Thouless1974}%
  \BibitemOpen
  \bibfield  {author} {\bibinfo {author} {\bibfnamefont {D.}~\bibnamefont
  {Thouless}},\ }\href {\doibase 10.1016/0370-1573(74)90029-5} {\bibfield
  {journal} {\bibinfo  {journal} {Phys. Rep.}\ }\textbf {\bibinfo {volume}
  {13}},\ \bibinfo {pages} {93} (\bibinfo {year} {1974})}\BibitemShut {NoStop}%
\bibitem [{Note3()}]{Note3}%
  \BibitemOpen
  \bibinfo {note} {Notice that the exponent describing the transport
  coefficient $\protect {\sim \delimiter "426830A \protect \qopname \relax
  o{exp}(-L/\xi )\delimiter "526930B }$ differs from the exponent $\protect
  {\nu =1}$ for the average correlation length, $\protect {\left \delimiter
  "426830A \xi \right \delimiter "526930B }$.\cite
  {BMSA:98,Mondragon-Shem:2013bs}}\BibitemShut {NoStop}%
\bibitem [{\citenamefont {Khmelnitskii}(1983)}]{Khmelnitskii83}%
  \BibitemOpen
  \bibfield  {author} {\bibinfo {author} {\bibfnamefont {D.~E.}\ \bibnamefont
  {Khmelnitskii}},\ }
\href {\doibase 10.1103/PhysRevB.55.1142} {\bibfield
  {journal} {\bibinfo  {journal} {JETP Lett. }\ }\textbf {\bibinfo
  {volume} {38}},\ \bibinfo {pages} {552} (\bibinfo {year}
  {1983})}
\BibitemShut {NoStop}%
\bibitem [{\citenamefont {Altland}\ and\ \citenamefont
  {Zirnbauer}(1997)}]{Altland1997}%
  \BibitemOpen
  \bibfield  {author} {\bibinfo {author} {\bibfnamefont {A.}~\bibnamefont
  {Altland}}\ and\ \bibinfo {author} {\bibfnamefont {M.~R.}\ \bibnamefont
  {Zirnbauer}},\ }\href {\doibase 10.1103/PhysRevB.55.1142} {\bibfield
  {journal} {\bibinfo  {journal} {Phys. Rev. B}\ }\textbf {\bibinfo
  {volume} {55}},\ \bibinfo {pages} {1142} (\bibinfo {year}
  {1997})}\BibitemShut {NoStop}%
\bibitem [{\citenamefont {Fulga}\ \emph {et~al.}(2011)\citenamefont {Fulga},
  \citenamefont {Hassler}, \citenamefont {Akhmerov},\ and\ \citenamefont
  {Beenakker}}]{Fulga2011}%
  \BibitemOpen
  \bibfield  {author} {\bibinfo {author} {\bibfnamefont {I.~C.}\ \bibnamefont
  {Fulga}}, \bibinfo {author} {\bibfnamefont {F.}~\bibnamefont {Hassler}},
  \bibinfo {author} {\bibfnamefont {A.~R.}\ \bibnamefont {Akhmerov}}, \ and\
  \bibinfo {author} {\bibfnamefont {C.~W.~J.}\ \bibnamefont {Beenakker}},\
  }\href {\doibase 10.1103/PhysRevB.83.155429} {\bibfield  {journal} {\bibinfo
  {journal} {Phys. Rev. B}\ }\textbf {\bibinfo {volume} {83}},\ \bibinfo
  {pages} {155429} (\bibinfo {year} {2011})}\BibitemShut {NoStop}%
\bibitem [{\citenamefont {Brouwer}\ \emph {et~al.}(1998)\citenamefont
  {Brouwer}, \citenamefont {Mudry}, \citenamefont {Simons},\ and\ \citenamefont
  {Altland}}]{BMSA:98}%
  \BibitemOpen
  \bibfield  {author} {\bibinfo {author} {\bibfnamefont {P.~W.}\ \bibnamefont
  {Brouwer}}, \bibinfo {author} {\bibfnamefont {C.}~\bibnamefont {Mudry}},
  \bibinfo {author} {\bibfnamefont {B.~D.}\ \bibnamefont {Simons}}, \ and\
  \bibinfo {author} {\bibfnamefont {A.}~\bibnamefont {Altland}},\ }\href@noop
  {} {\bibfield  {journal} {\bibinfo  {journal} {Phys. Rev. Lett.}\ }\textbf
  {\bibinfo {volume} {81}},\ \bibinfo {pages} {862} (\bibinfo {year}
  {1998})}\BibitemShut {NoStop}%
\end{thebibliography}
\end{document}